\documentclass{eptcs}
\providecommand{\event}{WLP 2016} 
\usepackage{breakurl}

\usepackage[english]{babel}
\usepackage[autostyle=true]{csquotes}

\usepackage{amsthm,amssymb}
\usepackage{mathtools}
\usepackage{graphicx}
\usepackage{subcaption}
\usepackage{listingsutf8}

\usepackage{booktabs}

\usepackage{tikz}
\usetikzlibrary{shapes.multipart}

\usepackage[boxed,noline]{algorithm2e}
\providecommand{\DontPrintSemicolon}{\dontprintsemicolon}
\SetAlCapSkip{\abovecaptionskip}
\SetAlCapSty{}

\usepackage{hyperref}
\usepackage[capitalise]{cleveref}

\newcommand{\ignore}[1]{} 

\newcommand{\prob}{{\sc ProB}}
\newcommand{\clpfd}{CLP(FD)} 
\newcommand{\clpfdbased}{CLP(FD)-based} 
\newcommand{\constraintbased}{constraint-based} 
\newcommand{\dpllt}{DPLL(T)} 
\newcommand{\smtlib}{SMT-LIB}

\newtheorem{definition}{Definition}[section]

\title{Constraint Logic Programming over Infinite Domains\\with an Application to Proof}
\author{
Sebastian Krings \qquad\qquad Michael Leuschel
\institute{
 Institut f\"{u}r Informatik, Heinrich-Heine Universit\"{a}t D\"{u}sseldorf\\
  Universit\"{a}tsstr. 1, D-40225 D\"{u}sseldorf\\
}
\email{\{krings,leuschel\}@cs.uni-duesseldorf.de}
}

\def\titlerunning{Constraint Logic Programming over Infinite Domains}
\def\authorrunning{S. Krings \& M. Leuschel}

\begin{document}
\maketitle

\begin{abstract}
We present a \clpfdbased\ constraint solver able to deal with unbounded domains.
It is based on constraint propagation, resorting to enumeration if all other methods fail.
An important aspect is detecting when enumeration was complete and if this has an impact on the soundness of the result.
We present a technique which guarantees soundness in the following way: if the constraint solver finds a solution it is guaranteed to be correct; if the constraint solver fails to find a solution it can either return the result ``definitely false'' in case it knows enumeration was exhaustive, or ``unknown'' in case it was aborted.
The technique can deal with nested universal and existential quantifiers.
It can easily be extended to set comprehensions and other operators introducing new quantified variables.
We show applications in data validation and proof.
\end{abstract}

\section{Introduction}
\label{sec:introduction}
Tool support is vital for the success of formal methods in general, and state-based formal methods in particular.
Some key technologies which enable effective validation of formal models are model checking, proof but \emph{constraint solving} as well.
Indeed, constraint solving enables animation and model checking of high-level formal models, it enables a large range of validation tasks
 ranging from bounded model checking to \constraintbased\ deadlock detection.
It is also crucial for test-case generation from formal models and can be used for finding counter examples to proof obligations.

Various techniques can be used to solve constraints expressed in specification languages like B, Z, TLA, Alloy, or VDM.\@
The main techniques are SAT solving, SMT solving and constraint programming.
SAT solving in this area has been made popular and practical via Alloy~\cite{Jackson:AlloyBook}.

Thus far, these techniques were often limited to first-order constraints and unable to deal with unbounded data values.
Let us examine the simple constraint $x=10*10$ over the integer variable $x$.
To solve this constraint using SAT solving the integer $x$ has to be represented by propositions, i.e., as a bit-vector.
For this, it is important to know how many bits are needed, both for $x$ itself and for intermediate values that can occur while
computing $x$. If one chooses too few bits, then the SAT solver may fail to find a solution where one exists, or report a solution
 where none exists (in case overflows are not detected).\footnote{Alloy recently has added overflow detection; but in case no model was found we do not know whether an overflow may have prevented
 finding a solution.}
Also, the encoding of values as bit vectors reaches its limits for more involved types like relations over large domains or higher-order values.
Take for example a relation $r$ which takes a set of elements over a domain $D$ and another set of elements over $D$.
If $D$ is of size 10, we need $2^{20}=$ bits to represent a possible value of $r$.
If $D$ is of size 20, we need $2^{40}=1,099,511,627,776$ bits, i.e., already 128 Gigabyte to store one variable.

\section{Constraint Solving}
\label{sec:constraintsolving}
The key challenges when solving constraints in high-level languages such as B
are universal and existential quantifications as well as set comprehensions and lambdas.
Each can be arbitrarily nested and are not limited to finite values.
So far we have tried different approaches to constraint solving as extensions to the \clpfdbased\ kernel of our model checker \prob\ in order to enable it to handle infinite domains:
We developed a translation to SAT via Kodkod~\cite{PlaggeLeuschel_Kodkod2012} and we integrated Z3~\cite{z3,probz3integration}.

The different techniques each have their own strengths and weaknesses:
While there are highly efficient algorithms for SAT solving, encoding of higher-order constraints is often not feasible.
Usually, the domain of integers has to be limited in order to allow bitlevel encoding of arithmetic.
This is even more problematic if higher-order logic or set theory come into play.
Due to a combinatorial blowup, resulting SAT constraints contain too many variables and become unsolvable.

SMT solvers on the other hand rely on decision procedures for different underlying logics.
Following the \dpllt\ algorithm, these solvers enumerate predicate values and try to infer logical consequences.
Hence, it is easy to extract a proof from an unsatisfiable query while it is difficult to extract a model out of a satisfiable one.

In contrast, \clpfd\ systems use constraint propagation and are more focused on data rather than predicates.
Indeed, they show the opposite behavior: a satisfiable query always returns a model.
At the same time it is difficult to extract proof of unsatisfiability.

\section{Technique}
\label{sec:technique}
In the following sections we describe how we extended \clpfd\ to enable handling of infinite domains and quantifiers.
In \cref{sub:enumwarning}, we will explain how we track enumeration of \clpfd\ variables.
Afterwards, we introduce a way to randomize the enumeration of large intervals in \cref{sub:randomenumeration}.
For simplicity, we will discuss our techniques on a small interpreter supporting only integer variables with arbitrary and possibly infinite domains together with
arithmetic expression on them, negation and nested existential and universal quantification.

\subsection{Detection and Categorization of Enumeration}
\label{sub:enumwarning}
It is common for constraint satisfaction problems, that domain propagation alone is not enough to infer values for participating variables.
This might be due to an underspecified problem or limitations in constraint solvers, e.g., global constraints that are too expensive to check.
Usually, constraint solvers rely on enumeration of possible values if all other methods fail.
However, there are some key difficulties:
\begin{enumerate}
\item Enumerating all values is only possible for reasonably sized domains.
Even for finite but large domains, exhaustive enumeration can be impossible due to computational limitations.
Obviously the same holds for infinite domains.
\item Depending on the scope of negations, quantifiers and other (nested) constructs finding a solution for a variable might imply both satisfiability and unsatisfiability of a (sub-)constraint.
\item Hence, if a solution is found it is not immediately clear if enumeration can be stopped.
\end{enumerate}

We intend to overcome these limitations by tracking the scope of enumerations, i.e., tracking in which contexts enumeration occur.
We distinguish the following types of enumerations, based on the effect on the overall solver result:
\begin{itemize}
\item Enumeration does not occur.
The result is not influenced, e.g., when no valuation is found the formula is unsatisfiable.
\item Enumeration is exhaustive.
In this case, all possible values for a variable were considered.
If no valuation is found, the formula in question is unsatisfiable.
\item Enumeration occurs and is not exhaustive.
In this case, we cannot directly infer if the formula is satisfiable or not and have to examine the context (aka scope)
 in which the enumeration occurred.
\end{itemize}

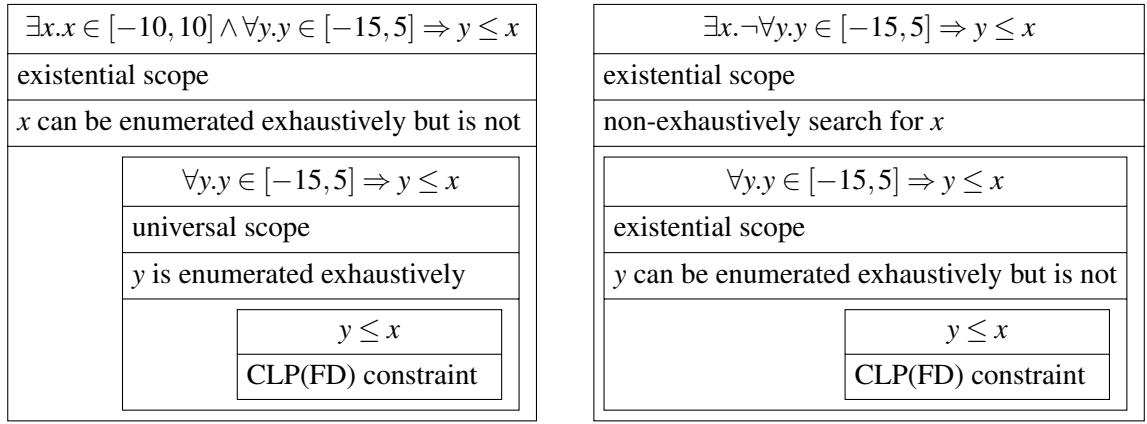
\begin{figure}
\begin{subfigure}[b]{0.49\textwidth}
\centering
\begin{tikzpicture}[
  double/.style={draw, anchor=text,
                 rectangle split,
                 rectangle split parts=2,
                 rectangle split part align={center,left}},
  quad/.style={draw, anchor=text,
                 rectangle split,
                 rectangle split parts=4,
                 rectangle split part align={center,left,left,right}}
  ]
  \node[quad,align=center,minimum width=200] {
      $\exists x . x \in [-10,10] \wedge \forall y . y \in [-15,5] \Rightarrow y \leq x$
    \nodepart{second}
      existential scope
    \nodepart{third}
      $x$ can be enumerated exhaustively but is not
    \nodepart{fourth}
      \tikz{\node[quad,minimum width=150] {
          $\forall y . y \in [-15,5] \Rightarrow y \leq x$
        \nodepart{second}
          universal scope
        \nodepart{third}
          $y$ is enumerated exhaustively
        \nodepart{fourth}
        \tikz{\node[double,minimum width=100] {
            $y \leq x$
          \nodepart{second}
            \clpfd\ constraint
          };
          }
        };
    }
  };
\end{tikzpicture}
\end{subfigure}
\begin{subfigure}[b]{0.49\textwidth}
    \centering
    \begin{tikzpicture}[
        double/.style={draw, anchor=text,
                       rectangle split,
                       rectangle split parts=2,
                       rectangle split part align={center,left}},
        quad/.style={draw, anchor=text,
                     rectangle split,
                     rectangle split parts=4,
                     rectangle split part align={center,left,left,right}}
    ]
      \node[quad,align=center,minimum width=200] {
          $\exists x . \neg \forall y . y \in [-15,5] \Rightarrow y \leq x$
        \nodepart{second}
          existential scope
        \nodepart{third}
          non-exhaustively search for $x$
        \nodepart{fourth}
          \tikz{\node[quad,minimum width=150] {
              $\forall y . y \in [-15,5] \Rightarrow y \leq x$
            \nodepart{second}
              existential scope
            \nodepart{third}
              $y$ can be enumerated exhaustively but is not
            \nodepart{fourth}
            \tikz{\node[double,minimum width=100] {
                $y \leq x$
              \nodepart{second}
                \clpfd\ constraint
              };
              }
            };
        }
      };
    \end{tikzpicture}
\end{subfigure}
\caption{Nested Enumeration Scopes}
\label{fig:nestedscopes}
\end{figure}

\Cref{fig:nestedscopes} shows the nesting of enumeration scopes for two simple predicates.
The outer variable $x$ is quantified existentially in both cases.
In the first one, we can enumerate all possible values exhaustively, while in the second case only non-exhaustive search is possible.
However, we only need to find one solution anyway, partial enumeration is not a problem.
The inner variable $y$ can be enumerated exhaustively.
In the first example, we have to do so in order to validate the universal quantification.
In the second example, exhaustive enumeration is possible but not necessary.

For further examples, take a look at the following constraints:
\begin{itemize}
\item \texttt{x*x = 10000} is true,
a solution (\texttt{x = -100}) can be computed without resorting to enumeration.
\item Using backtracking, we can find all solutions.
Again, no enumeration is needed to compute
\texttt{\{x|x*x=10000\} = \{-100,100\}}.
\item In contrast, enumeration is necessary to find a solution to
\texttt{x>10000 \& x mod 1234 = 1} as propagation of \texttt{mod} does not render the domain of \texttt{x} finite.
However, the solution found is sound, enumeration did not influence the result.
\item As a consequence, we cannot compute all solutions.
Our approach is unable to solve \texttt{\{x|x>10000 \& x mod 1234 = 1\}}.
\item For certain unsatisfiable constraints, such as \texttt{x*x = 10001}, \clpfd\ detects unsatisfiability by domain propagation.
No enumeration occurs, the predicate is guaranteed to be false.
\item In contrast, no solution is found for \texttt{x>10000 \& x mod 1234 = 1 \& x*x = 10*x}.
Common propagation rules such as the ones used in SWI Prolog's \clpfd~\cite{Triska2014,Apt2007} are too weak to conclude $x = 0 \vee x = 10$ from $x*x = 10*x$ as long as there is no upper bound attached to $x$.
In consequence, we cannot detect unsatisfiability as we have to (partially) enumerate the  infinite domain of \texttt{x}.
Since no solution has been found, enumeration cannot be ignored;
Indeed, the predicate might still be true.
\end{itemize}

\subsubsection{Setting up Constraints}
Constraints are set up using two Prolog predicates, \texttt{solve} for positive and \texttt{solve\_not} for negative constraints.
\clpfd\ constraints are immediately set up. Part of the interpreter is shown in \cref{lst:core}, the complete
source code and usage instructions can be obtained from \url{https://github.com/wysiib/infinite_domain_solver}.

\begin{lstlisting}[frame=single,float=t,firstline=1,lastline=56,label=lst:core,abovecaptionskip=\abovecaptionskip,captionpos=b,caption=Core of Interpreter]
solve_constraint(Constraint,TopLevelVars) :-
    retractall(enum_warning),
    solve(Constraint,ExistsWF,AllWF),
    ground_vars(TopLevelVars,ExistsWF,AllWF).

ground_vars(TopLevelVars,ExistsWF,AllWF) :-
    maplist(enumerate_exists_aux,TopLevelVars),
    ground_wfs(ExistsWF,AllWF).

solve(A & B,EWF,AWF) :- solve(A,EWF,AWF), solve(B,EWF,AWF).
solve(A or B,EWF,AWF) :- solve(A,EWF,AWF) ; solve(B,EWF,AWF).
...
solve(not(A),EWF,AWF) :- solve_not(A,EWF,AWF).
solve(V in D,_,_) :- V in D.
solve(A = B,_,_) :-
    compute_exprs(A,B,AE,BE),
    AE #= BE.
...
solve(forall(X,LHS => RHS),_EWF,AWF) :-
    when(ground(AWF),enumerate_forall(X,LHS,RHS)).
solve(exists(X,RHS),EWF,_AWF) :-
    when(ground(EWF),enumerate_exists(X,RHS)).

solve_not(A & B,EWF,AWF) :-
    solve_not(A,EWF,AWF) ; solve_not(B,EWF,AWF).
solve_not(A or B,EWF,AWF) :-
    solve_not(A,EWF,AWF), solve_not(B,EWF,AWF).
...
\end{lstlisting}

To control enumeration, we pass around two ``wait flags''.
The first is used to trigger setup of constraints and enumeration of variables that are existentially quantified, i.e., they are introduced by $\exists$ or $\neg \forall$.
The other does the same for universal quantification.

\Cref{lst:setupquantifiers} shows the implementation in our simple solver.
Once a wait flag is grounded, the code in \cref{lst:enumforall,lst:enumexists} is called:
We set up the inner constraints of the quantifier using fresh wait flags, that will later be grounded as well.
This accounts for a hierarchy of scopes, where each may need to find a single solution or inspect all possible solutions for a variable.
Again, see \cref{fig:nestedscopes} for an example.

\begin{lstlisting}[frame=single,float=t,firstline=53,lastline=56,label=lst:setupquantifiers,abovecaptionskip=\abovecaptionskip,captionpos=b,caption=Setup of Quantifiers]
solve(forall(X,LHS => RHS),_EWF,AWF) :-
    when(ground(AWF),enumerate_forall(X,LHS,RHS)).
solve(exists(X,RHS),EWF,_AWF) :-
    when(ground(EWF),enumerate_exists(X,RHS)).
\end{lstlisting}

\subsubsection{Enumeration}
Enumeration occurs in two phases.
First, we ground the wait flag triggering enumeration of existentially quantified variables.
This leads to the coroutines set up in \cref{lst:setupquantifiers} being resumed.
The key difference between enumerating an existentially quantified variable (\cref{lst:enumexists}) and a universally quantified one (\cref{lst:enumforall}) lies within the solver's reaction to infinite domains.
Existentially quantified ones are enumerated as shown in \cref{lst:enumexists}:
\begin{enumerate}
\item The inner constraint of the quantifier is set up as usual,
\item Inner variables are enumerated,
\item We begin enumerating the quantified variable:
\begin{itemize}
\item If the domain is infinite, we store that enumeration cannot be exhaustive. Afterwards, we start enumerating.
\item Otherwise, we use regular \clpfd\ labeling.
\end{itemize}
\end{enumerate}

\begin{lstlisting}[frame=single,float=t,firstline=103,lastline=117,label=lst:enumexists,abovecaptionskip=\abovecaptionskip,captionpos=b,caption=Enumerate Existentially Quantified Variable]
enumerate_exists(Var,RHS) :-
    % setup inner constraints
    solve(RHS,NewEWF,NewAWF), !,
    ground_wfs(NewEWF,NewAWF),
    enumerate_exists_aux(Var).
enumerate_exists_aux(Var) :-
    fd_size(Var,sup), !,
    % non-exhaustively enumerate infinite domain
    % need to find just one element!
    assert(enum_warning),
    fd_inf(Var,Min), fd_sup(Var,Max),
    enumerate_infinite(Var,0,Min,Max).
enumerate_exists_aux(Var) :-
    indomain(Var).
\end{lstlisting}

In the second step, we cannot enumerate all possible values of the variable in question.
We thus have to decide on a value selection strategy.
Our simple example interpreter enumerates as follows: Starting from zero we alternate between the positive and negative value of an increasing counter, skipping values not in the corresponding domains. As soon as both upper and lower bounds are passed, the domain has been enumerated exhaustively.

This simple enumeration pattern is not sophisticated enough for complicated constraints.
To improve, one could rely on techniques like the level diagonalization suggested in~\cite{Christiansen2008}.

In \prob, we combine the simple enumerator with other (non-exclusive) strategies like case splits.
Another alternative is to use random enumeration as we will show in \cref{sub:randomenumeration}.
After all existentially quantified variables have been enumerated, we ground the wait flag that triggers universal quantifiers.
We enumerate as outlined in \cref{lst:enumforall}:
\begin{enumerate}
\item The inner constraint of the quantifier is set up as usual,
\item We begin enumerating the variable:
\begin{itemize}
\item If the domain is infinite, we throw an enumeration exception.
We would need to fully enumerate an infinite domain in order to solve the constraint.
This futile attempt is dropped.
Keep in mind, that the occurrence of infinite domains is usually influenced by the order of variables.
By enumerating variables with finite domains first, we might shrink other domains.
\item Otherwise, we try all values in its domain to check the universal quantification.
\end{itemize}
\end{enumerate}

With this extension, our \clpfdbased\ solver is able to handle both existential and universal quantification.
In several cases, for instance in $y=2 \wedge \forall x.(x \in [0,10] \Rightarrow x>y)$ the solver can recognize exhaustiveness of labelings.
As a result, satisfiability and unsatisfiability can be deduced and reported to the user.
In case of infinite or large domains the solver can tell if a result is still valid, despite the fact that a domain has not been enumerated completely.
To increase coverage of large domains in the face of timeouts, the following section will introduce random enumeration.

\begin{lstlisting}[frame=single,float=t,firstline=82,lastline=98,label=lst:enumforall,abovecaptionskip=\abovecaptionskip,captionpos=b,caption=Enumerate Universally Quantified Variable]
enumerate_forall(Var,LHS,RHS) :-
    LHS = (_ in Min .. Max),
    % setup of inner constraints: there is a choicepoint here
    % to allow for different solutions to inner variables
    solve(LHS & RHS,NewEWF,NewAWF), !,
    ground_wfs(NewEWF,NewAWF),
    enumerate_forall_aux(Min,Max,Var).
% need to exhaustively enumerate infinite domain -> exception
enumerate_forall_aux(_,sup,_) :- throw(enum_infinite).
enumerate_forall_aux(inf,_,_) :- throw(enum_infinite).
% domain is finite, try all elements
enumerate_forall_aux(Current,Max,Var) :-
    Current =< Max, !,
    try_forall_value(Current,Var), % does not bind variable
    Current2 is Current + 1,
    enumerate_forall_aux(Current2,Max,Var).
enumerate_forall_aux(_,_,_).
\end{lstlisting}

\subsection{Randomized Enumeration of Large Intervals}
\label{sub:randomenumeration}
Another limitation of most \clpfd\ systems lies in how the next value of a variable is selected upon labeling.
Within SICStus, the user can select between the following strategies~\cite{sicstusmanual} together with the options \emph{up}, to use ascending order and \emph{down}, to use descending order:
\begin{itemize}
\item \emph{step}, i.e., a binary choice between $X = B$ and $X \neq B$, where B is the lower or upper bound of X.
\item \emph{enum}, i.e., multiple choice for X corresponding to the values in its domain.
\item \emph{bisect}, i.e., binary choice between $X\leq M$ and $X > M$, where $M = \lfloor \frac{\min(X) + \max(X)}{2} \rfloor$.
\item \emph{median}, i.e., binary choice between $X = M$ and $X \neq M$, where M is the median of the domain.
\item \emph{middle}, i.e., binary choice between $X = M$ and $X \neq M$, where $M = \lfloor \frac{\min(X) + \max(X)}{2} \rfloor$.
\end{itemize}

Summarizing, \clpfd\ variables can be set to domain values in various ways using domain splitting or simple enumeration.
One property is common to all strategies.
The domains are traversed deterministically.
In case the domains are small enough, i.e., they can be enumerated exhaustively, there is no need for a different strategy.
However, for large domains this might not be sufficient.
Instead of traversing the domain linearly, it could be beneficial to use a random permutation:
\begin{itemize}
\item For large domains, values of different sizes will be tried out before a timeout occurs.
\item It is less likely to get stuck in some part of the search space where there is no solution.
If we fear search is stuck we could restart as described in~\cite{randomrestarts}.
This is common in SAT and SMT solvers.
\item For applications like test case generation it is desirable to compute test inputs that fulfill some coverage criterion, e.g., that certain intervals or sets of parameters or values have been used.
With linear enumeration and backtracking, generated test cases might only differ in the variable set last.
\end{itemize}

Note that we want to compute a random permutation of the domain.
To avoid duplicates we do not want to randomly draw elements from the domain.
Furthermore, we have to keep track of the exhaustiveness of our enumeration in order to detect unsatisfiability.

A classic algorithm to compute random permutations for given intervals is the Fisher-Yates shuffle~\cite{fisheryates} or Knuth shuffle~\cite{Knuth:art2}.
Its pseudo code can be found in \cref{alg:fisheryates}.
The algorithm has a weakness: the list to be shuffled has to be in memory completely.
This is not feasible for intervals too large to be stored.
We hence need an algorithm allowing us to compute a random permutation on the fly.

\begin{algorithm}[t]
 \DontPrintSemicolon{}
 \KwData{List $a$}
 \KwResult{Random permutation of $a$}
 \For{$i \in [0, length(a) - 1]  $} {
 chose $j$ randomly such that $0 \leq j \leq i$ \;
 \If{$j \neq i$}{$perm[i] \coloneqq\ perm[j]$}
 $perm[j] \coloneqq\ a[i]$
 }
 \Return\ $perm$
 \caption{Fisher-Yates / Knuth Shuffle}
\label{alg:fisheryates}
\end{algorithm}

One such algorithm can be constructed using cryptographic techniques as outlined in~\cite{randomperms}.
The key idea is to construct an encryption function encrypting the elements to be permuted onto themselves.
In order to allow for later decryption, an encryption function has to be unambiguously, i.e., given a fixed key there has to be a one-to-one mapping between plaintext and ciphertext.
This will ensure, that we do in fact compute a permutation, i.e., we do not add or remove elements.
Before we can present our implementation, we introduce a few definitions regarding ciphers.
The definitions are following~\cite{Menezes:1996:HAC:548089}.

\begin{definition}[Block Cipher]
An $n$-bit \emph{block cipher} is a function $E : {[0,1]}^n \times K \rightarrow {[0,1]}^n$, such that for each key $K \in \mathcal{K}$, $E(P,K)$ is an invertible mapping from ${[0,1]}^n$ to ${[0,1]}^n$, written $E_K(P)$.
$E_K(P)$ is called the \emph{encryption function} for $K$.
 The inverse mapping is the \emph{decryption function}, denoted $D_K(C)$.
 $C = E_K(P)$ denotes that ciphertext $C$ results from encrypting plaintext $P$ under $K$.
\end{definition}

\begin{definition}[Random Cipher]
A (true) \emph{random cipher} is an $n$-bit block cipher implementing all $2n!$ bijections on $2n$ elements. Each of the $2n!$ keys specifies one such permutation.
Obviously, the key space for a true random cipher is too large to be used in practice.
\end{definition}

\begin{definition}[Iterated Block Cipher]\label{def:iteratedcipher}
An \emph{iterated block cipher} is a block cipher involving the sequential repetition of an internal function called a round function.
Parameters include the number of rounds $r$, the block bit size $n$, and the bit size $k$ of the input key $K$ from which $r$ subkeys $K_i$, one for each round, are derived.
For invertibility, for each $K_i$ the round function is a bijection on the round input.
\end{definition}

\begin{figure}[t]
\centering
\includegraphics[width=\textwidth]{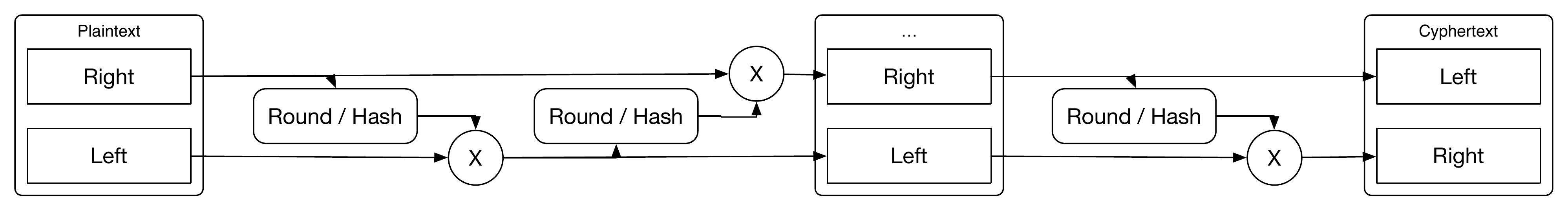}
\caption{Feistel Network}
\label{fig:feistelnetwork}
\end{figure}

\begin{definition}[Feistel Cipher]\label{def:feistelcipher}
A \emph{Feistel cipher} is an iterated block cipher mapping a $2t$-bit plaintext $(L_0, R_0)$, for t-bit blocks $L_0$ and $R_0$, to a ciphertext $(R_r,L_r)$, through an $r$-round process where $r \geq 1$.
For $1 \leq i \leq r$, round $i$ maps $(L_{i-1},R_{i-1})$ to $(L_i,R_i)$ as follows:
\begin{equation*}
L_i = R_{i-1}, R_i = L_{i-1} \oplus f(R_{i-1}, K_1).
\end{equation*}
For each round, the \textit{subkey} $K_i$ is derived from the given cipher key $K$.
$f$ is the \textit{round function} mentioned in \cref{def:iteratedcipher}.
Often other (product) ciphers are used as $f$.
However, F does not need to be invertible for the Feistel cipher to be.
\Cref{fig:feistelnetwork} shows how a Feistel cipher operates on the input blocks.
\end{definition}

Now, we can follow one of the approaches suggested by~\cite{cipherswithfinitedomains} to construct a cipher that does not operate on ${[0,1]}^n$ but rather on $[0,k]$.
The authors describe the construction of a generalized Feistel cipher:
\begin{enumerate}
\item Choose $a,b \in \mathbb{N}$ with $ab \leq k$ to decompose $m$ taken from $[0,k]$ into $L = m \mod a$ and $R = \lfloor \frac{m}{a} \rfloor$.
\item Use $(L,R)$ as inputs to a Feistel cipher as defined in \cref{def:feistelcipher}.
\item Perform a given number of iterations using random round functions whose ranges contain $[0,k]$.
\end{enumerate}

Observe that \cref{def:feistelcipher} assumes that $L$ and $R$ have the same length.
Hence, the overall bit-length has to be divisible by 2\footnote{For technical reasons the actual implementation uses a bit-length divisible by 4.}.
With the construction above, we can thus only create ciphers for certain intervals, i.e., we can construct a cipher $[0,15] \rightarrow [0,15]$ but not $[0,5] \rightarrow [0,5]$.
However, the cipher $[0,15] \rightarrow [0,15]$ can be used to permute $[0,5]$ by iterating to the next index if a number $\leq 6$ is drawn.
Intervals that do not start with 0 are shifted. The shift is later re-added to the drawn number.

The implementation of random permutations is outlined in \cref{alg:randompermutation} and \cref{alg:randompermutationnext}.
We can use a simplified version of the construction in~\cite{cipherswithfinitedomains} because we do not rely on strong cryptographic properties.
On the interval $[1,6]$ it proceeds as follows:
\begin{enumerate}
\item Compute the interval length $l = 6-1+1$.
\item Find the number of bits need to store $l$. In this case $l = 6$ means we need at least 3 bits.
\item Round up to the next even number to allow symmetric split into $L$ and $R$.
We can split an interval of 4 bits. Hence, we need to take into account all inputs from $[0,15]$.
\item Compute the bit masks $BL = 1100$ and $BR = 0011$ used to extract the first 2 and the last 2 bits.
\item Set the current index to 0 and select a random key to choose a permutation.
\item To draw the next element of the random permutation:
\begin{enumerate}
\item Use a Feistel cipher to encrypt the current index using the key.
\item Increment the index.
\item Repeat if the cipher value is larger than 5, else return $(5+1)$ to stay within range.
\end{enumerate}
\end{enumerate}

\begin{algorithm}[t]
 \DontPrintSemicolon{}
 \KwData{Interval $I$ to draw from}
 \KwResult{Bitmasks $BL$ to extract $L$, $BR$ to extract $R$, number of bits $n$}
$length = \max(I) - \min(I) + 1$ \;
$n = \ln(length)$ \;
\If{$n$ is odd}{
$n = n + 1$
}
$BR = 2^{\frac{n}{2} - 1}$ \;
$BL = 2^{n} - 1 - BR$
\caption{Random Permutation: Setup}
\label{alg:randompermutation}
\end{algorithm}

\begin{algorithm}[t]
\SetKwRepeat{Do}{do}{while}
 \DontPrintSemicolon{}
 \KwData{Interval $I$, current index $c$, max index $maxIdx$, left/right mask $BL,BR$, number of bits $n$}
 \KwResult{Next random element $rnd$ and next index to draw $next$}
\Do{$rnd > \max(I) - \min(I) \wedge c < \max$}{
$left = c \mathrel{\&} BL >> \lfloor \frac{n}{2} \rfloor$ \;
$left = c \mathrel{\&} BR$ \;
$left, right = feistel\_rounds(left,right)$ \;
$rnd = (left << \lfloor \frac{n}{2} \rfloor) {\mathrel{|}} right$ \;
$c = c + 1$
}
\eIf{$c > maxIdx$} {
\Return\ failure, permutation enumerated exhaustively
}{
\Return\ $rnd$, $next = c + 1$
}
\caption{Random Permutation: Next Element}
\label{alg:randompermutationnext}
\end{algorithm}

For the rounding function one can use a hash function such as Prolog's term\_hash.
We use a given random seed as the encryption key.
All subkeys are identical to the main key.

\section{High-Level Reasoning using CHR}
\label{sub:chr}
So far, \prob\ was mostly used for animation, model checking and data validation~\cite{2014formal}.
Hence, it used to be tailored towards finding solutions to satisfiable formulas, which is where constraint programming has its strengths.
Therefore, choosing \clpfd\ as a basis for \prob\ has been a reasonable decision:
\begin{itemize}
\item It can deal with large and only partially known data.
\item Even though B is a higher-order language including sets, relations and functions, everything could be expressed by or in conjunction with \clpfd\ variables and constraints.
\item Support of reification made it considerably easier to propagate information between the different solvers, i.e., from integer constraints to set constraints and vice versa.
\item Although satisfiability of predicates written in B is in general undecidable, \prob\ is able to solve a useful class of subproblems.
\end{itemize}

New applications like symbolic model checking however made shortcomings of \clpfd\ obvious:
Even for simple constraints like $x < y \wedge y < x$
contradictions often can only be detected if variables have finite domains.
Again, this is due to propagation relying on finite upper and lower bounds~\cite{Triska2014,Apt2007}.

To improve, we implemented a set of rules working on top of the \clpfd\ variables.
Our initial idea was to perform some kind of high-level propagation, where new \clpfd\ constraints are discovered from the existing constraints.
That is, we would implement propagation rules like the transitivity of $<$ stating that $x < y \wedge y < z \Rightarrow x < z$.
These rules are implemented in CHR~\cite{Fruehwirth199895,frühwirth2009constraint}, a committed choice language that can be embedded in Prolog.
CHR supports three different kinds of rules to modify a \emph{constraint store} holding the current state of constraints:
\begin{itemize}
\item Simplification rules of the form $h_1, \dots, h_n \,|\, g_1, \dots, g_m \Longleftrightarrow b_1, \dots, b_o$. Here, $h_1, \dots, h_n$ are the so called \enquote{head}, i.e., constraints that have to be found in the constraint store for the rule to act upon. $g_1, \dots, g_m$ are called \enquote{guards}. These are predicates that have to hold for the rule to be allowed to fire. If the rule can be executed, the heads are rewritten into the  \enquote{bodies} $b_1, \dots, b_o$, i.e., $h_1, \dots, h_n$ are removed from the constraint store while $b_1, \dots, b_o$ are added.
\item Propagation rules of the form $h_1, \dots, h_n \,|\, g_1, \dots, g_m \Longrightarrow b_1, \dots, b_o$. Here, the bodies are added to the constraint store without removing the heads.
\item Simpagation rules such as $h_1, \dots, h_{l} \,\backslash h_{l+1}, \dots, h_n \,|\, g_1, \dots, g_m \Longleftrightarrow b_1, \dots, b_o$ combine the former. The constraints $h_1, \dots, h_{l}$ are kept in the constraint store while $h_{l+1}, \dots, h_n$ are removed.
\end{itemize}

We augmented the constraint solver with CHR rules handling integer arithmetic, focusing on detection of contradictions involving linear inequalities.
The rules are comparable to those introduced in the finite domain solver of~\cite[Ch. 8]{frühwirth2009constraint}.
However, we do not handle domains inside CHR, but rather integrate with \clpfd.
Regarding the implementation of infinite domain solvers in CHR see~\cite[Ch. 9]{frühwirth2009constraint}.

\Cref{lst:chrint} includes an extract of the CHR rules encoding properties of $<$ and $\leq$.
As can be seen, we introduced rules for (anti-)reflexivity, antisymmetry, idempotence and transitivity.
For instance, transitivity of $\leq$ is given by the CHR rule \texttt{leq(X,Y), leq(Y,Z) ==> leq(X,Z)}, stating that from $X \leq Y \wedge Y \leq Z$ we can infer $X \leq Z$.
Newly inferred constraints are submitted to the underlying \clpfd\ system.

In addition to inferring new constraints, CHR rules can be used to infer unsatisfiability.
As an example, look at the rule encoding the antireflexivity of $<$.
It states that if we have $X < X$, \texttt{fail} has to be executed.
Due to CHR being a committed-choice language, the whole CHR run is discarded.
The Prolog rule that added the offending constraint by calling \texttt{leq(X,X)} fails.
In consequence, further propagation is stopped and the failure has to handled by the solver.

Of course a fully fledged solver needs to include several other CHR rules dealing with common cases of integer constraints.
As can be seen in the examples in \cref{table:chrbenchmarks}, the example set of CHR rules is far from being complete:
It is able to handle simple cases like $ x > y \wedge y > x$ and transitive cases like $w>x \wedge x>y \wedge y>z \wedge z>w$.
However, it is so far unable to do simple arithmetic as in $x+2 > y+1 \wedge y > x$.
In \prob\ we have added both arithmetic as well as further high-level rules.

\begin{lstlisting}[frame=single,float=t,firstline=8,lastline=17,abovecaptionskip=\abovecaptionskip,captionpos=b,caption={CHR Rules for Integer Inequalities},label=lst:chrint]
reflexivity  @ leq(X,X) <=> true.
antisymmetry @ leq(X,Y), leq(Y,X) <=> X = Y.
idempotence  @ leq(X,Y) \ leq(X,Y) <=> true.
transitivity @ leq(X,Y), leq(Y,Z) ==> leq(X,Z).

antireflexivity @ lt(X,X) <=> fail.
idempotence     @ lt(X,Y) \ lt(X,Y) <=> true.
transitivity    @ lt(X,Y), leq(Y,Z) ==> lt(X,Z).
transitivity    @ leq(X,Y), lt(Y,Z) ==> lt(X,Z).
transitivity    @ lt(X,Y), lt(Y,Z)  ==> lt(X,Z).
\end{lstlisting}

\begin{table}[tbp]
\centering
\begin{tabular}{lcc}\toprule
predicate & with CHR & without CHR\\ \midrule
$x > 3$ & 1.2 & 1.08 \\
$x > y \wedge y > x$ & 0.92 & timeout \\
$x = 3 \wedge x > y \wedge y = 4$ & 0.98 & 1.07 \\
$x = 3 \wedge x > y$ & 0.86 & 1.0 \\
$x = 3 \wedge x < y$ & 1.1 & 1.11 \\
$w > x \wedge x > y \wedge y > z \wedge w = 1 \wedge z = 1$ & 0.98 & 0.95 \\
$w > x \wedge x > y \wedge y > z \wedge z > w$ & 0.88 & timeout \\
$x + 2 > y + 1 \wedge  y > x$ & timeout & timeout \\
$x > y \wedge y > x+1$ & 0.9 & timeout \\
 \bottomrule
\end{tabular}
\caption{Small Benchmarks with / without CHR (in s)}
\label{table:chrbenchmarks}
\end{table}

\section{Applications}
\label{sec:applications}
Within our model checker and animator \prob\ we have several applications for an infinite domain constraint solver. We already mentioned animation, model checking and \constraintbased\ validation in the introduction. In this section, we will look at other applications.

We start with an overview of \prob's kernel in \cref{sub:solvingkernel}.
In \cref{sub:disprover} we explain how our constraint solving technique can be used for proof and disproof of B proof obligations.
Furthermore, we used it to solve SMT problems.
As a second application, we present our effort to use \prob\ for data validation tasks. \Cref{sub:datavalidation} will describe the main challenges.

\subsection{The \prob\ Constraint Solving Kernel}
\label{sub:solvingkernel}
\begin{figure}
\begin{center}
  \includegraphics[width=7.9cm]{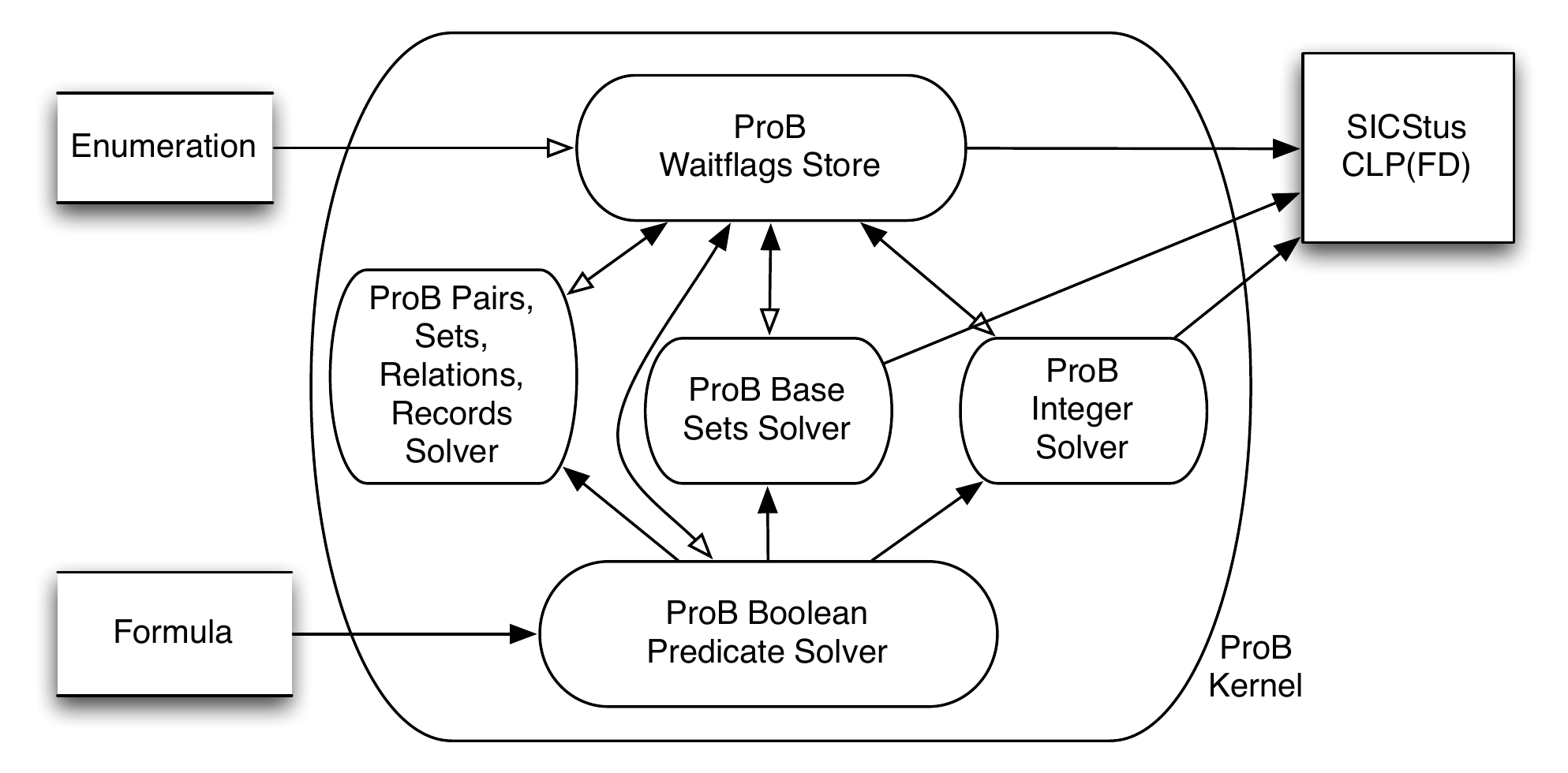}
  \caption{A View of the \prob\ Kernel}\label{ProB_CBC_Kernel}
\end{center}
\end{figure}

The \prob\ kernel can be viewed as a constraint-solver for the basic datatypes of B and the various operators on it.
It heavily relies on the SICStus Prolog \clpfd\ system~\cite{sicstusclpfd} which follows the general implementation scheme of~\cite{DBLP:conf/iclp/JaffarM87}.
It supports booleans, integers, user-defined base types, pairs, records and inductively: sets,
 relations, functions and sequences.
These datatypes and operations are embedded inside B predicates, which can make use of the
 usual logical connectives ($\wedge, \vee, \Rightarrow, \Leftrightarrow, \neg$) and typed universal ($\forall x. P \Rightarrow Q$)
  and existential ($\exists x . P \wedge Q$) quantification.
An overview of the various solvers residing within the \prob\ kernel can be seen in \cref{ProB_CBC_Kernel}.

Different solvers are linked via reification variables.
Enumeration is controlled by setting up coroutines at various choice points.
For more fine-grained enumeration, coroutines might be set up together with a wait flag that allows the solver to trigger delayed constraints.
The techniques presented in this paper are fully integrated with the solving kernel and its features.

\subsection{Prove and Disprove}
\label{sub:disprover}
As mentioned above, one of our key motivations is to move \prob\ from being guaranteed sound for finite domains to infinite domains. This is particularly important if \prob\ is to be used as a prover.

In~\cite{disprover,disprover_eval}, we embedded \prob\ into Rodin~\cite{rodinplatform}, an IDE for Event-B, in order to generate counter-examples for proof obligations.
Given a sequent with goal $G(x_1, \ldots, x_k)$ and hypotheses $H_i(x_1, \ldots, x_k)$ we build the predicate
\begin{equation*}
\exists x_1, \ldots, x_k : (H_1(x_1, \ldots, x_k) \wedge \ldots \wedge H_n(x_1, \ldots, x_k)) \Rightarrow \neg G(x_1, \ldots, x_k)
\end{equation*}
and feed it to our constraint solver. If the predicate does hold, \prob\ returns a valuation for $x_1, \ldots, x_k$, representing a counter-example to the sequent.

In general, checking the satisfiability of propositional formulas is NP complete.
Beyond that, typical Event-B proof obligations consist of first order logic formulas, for which the problem becomes undecidable.
Previously~\cite{prob2003,prob2008,disprover}, we overcame this limitation by limiting domains to be finite.
However, this prevents drawing any conclusions from the absence of a counter-example.

The techniques presented in this work have been used to extend the disprover to a fully fledged prover.
By observing the state of enumerations as explained in \cref{sub:enumwarning}, \prob\ is able to tell if the search for a counter-example was exhaustive.
If this is the case, we can report a proof to the user.

\begin{figure}[t]
\centering
\begin{subfigure}[b]{0.48\textwidth}
\includegraphics[width=\textwidth]{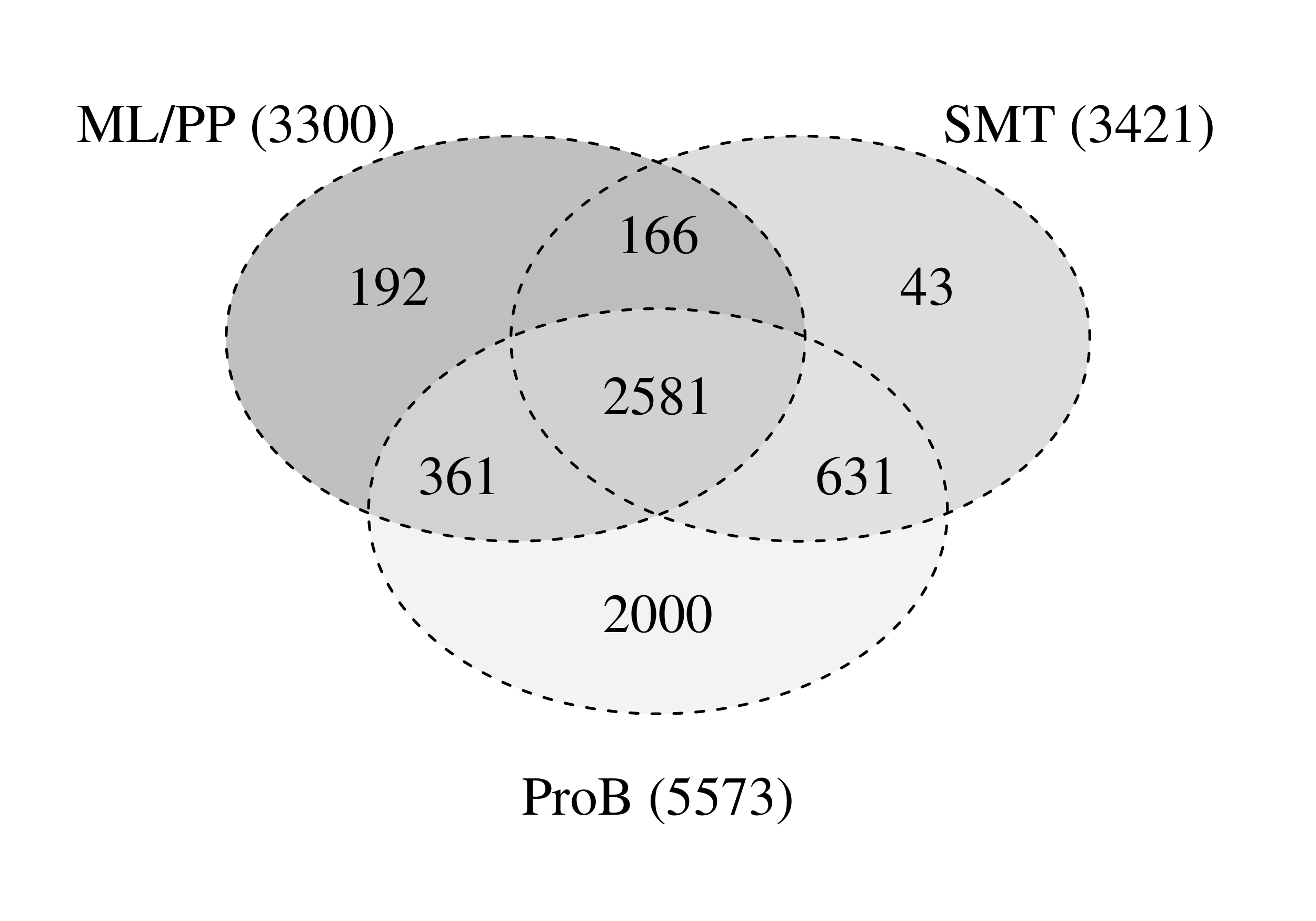}
\caption{(Dis-)Proof}
\label{fig:prover}
\end{subfigure}
\begin{subfigure}[b]{0.48\textwidth}
\includegraphics[width=\textwidth]{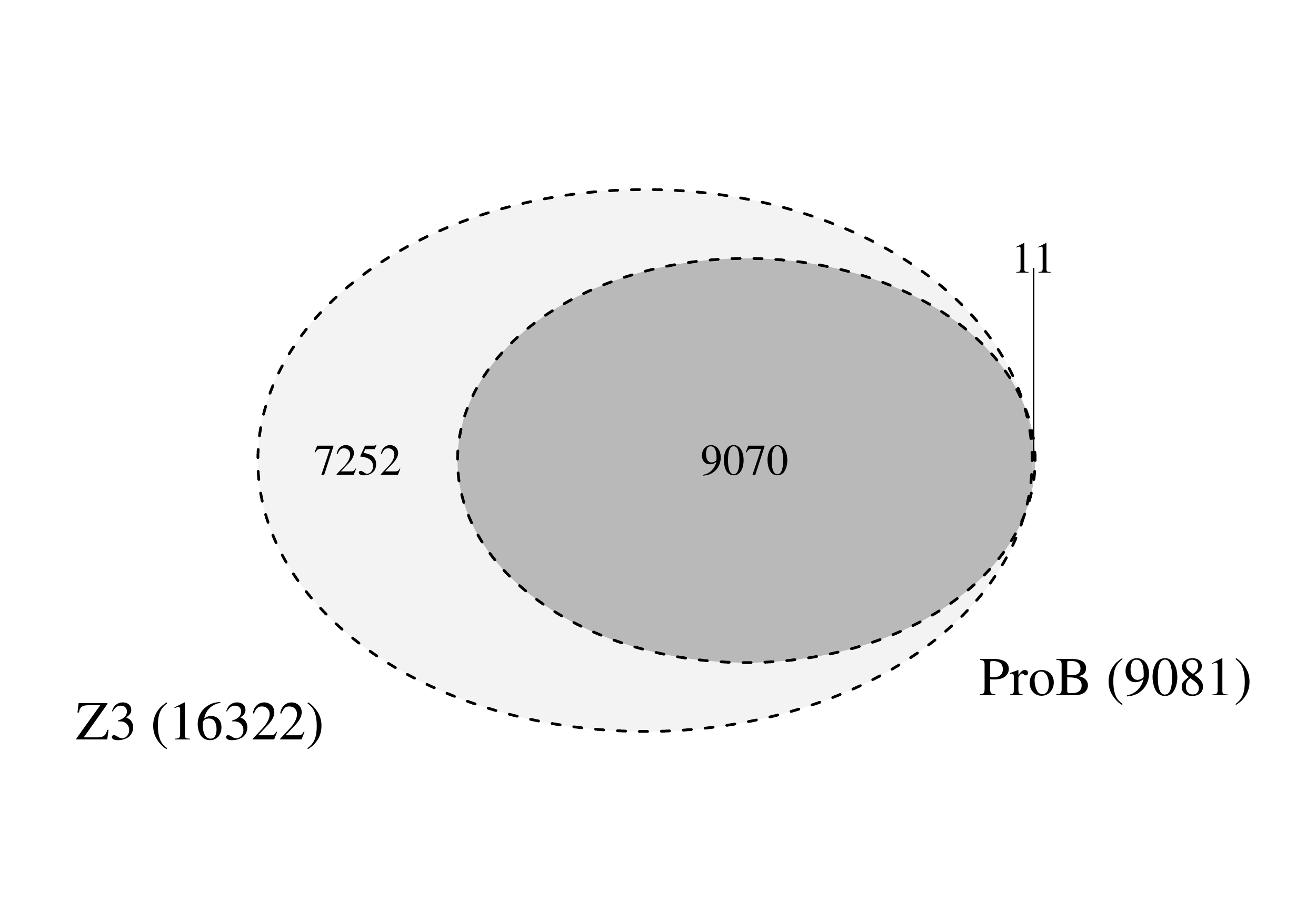}
\caption{SMT Solving}
\label{fig:smtsolver}
\end{subfigure}
\caption{Benchmarks}\label{fig:benchmarks}
\end{figure}

We performed several benchmarks comparing our prover to ML and PP~\cite{atelierb42}, two specialized provers for B and Event-B.
Additionally, we compared the performance of SMT solvers on the proof obligations.
An overview is given in \cref{fig:prover}, where we benchmarked on a diverse selection of different B and Event-B machines.
As you can see, \prob\ outperforms the other proves.
Details can be found in~\cite{disprover_eval}.

\subsubsection{SMT Solving}
The techniques presented can also be used to solve SMT problems, such as the ones collected by the \smtlib\ project.
In particular, we used all benchmarks that involve (non-)linear integer arithmetic and quantification but no other constructs like arrays or bit vectors.
The results can be seen in \cref{fig:smtsolver}.
\prob\ augmented with enumeration tracking cannot compete with Z3~\cite{z3}.
Yet, a considerable number of both satisfiable and unsatisfiable benchmarks can be solved using our technique.

\subsection{Data Validation}
\label{sub:datavalidation}
Model checking and \constraintbased\ validation aside, \prob\ and B are often used for data validation tasks~\cite{HansenSchneiderLeuschel:ABZ2016}.
The challenge here is to efficiently handle large relations and sets, e.g., representing track topologies,
 and at the same time effectively solving constraints and dealing with certain infinite functions which are used to manipulate data.
Notable applications come from the railway industry~\cite{FalampinLeuschelDeployBook} or university timetabling~\cite{timetable_validation}.

\section{Related Work}
SMT solvers~\cite{BSST09} such as Z3~\cite{z3} are able to handle infinite domains and quantifiers.
As outlined in the introduction, a major difference lies within the handling of data.
While SMT solvers are more focused on predicates, \clpfd\ systems are more oriented towards data.
As a result, SMT solvers can handle infinite domains more efficiently and detect unsatisfiability in more cases.
However, model generation (in particular in the presence of quantifiers) is often easier using \clpfd.

In~\cite{probz3integration} we investigated a different approach to add high-level reasoning to a \clpfdbased\ solver.
Instead of implementing rules in CHR we connected the SMT solver Z3 to SICStus Prolog.
We transferred each predicate asserted in \clpfd\ to Z3 and queried both solvers for a solution.
Furthermore, intermediate assignments and domains could be communicated back and forth.

While the approach was more powerful when it comes to reasoning, using CHR rules has the advantage of an immediate integration into Prolog.
Hence, there is no communication overhead and no translation between different representations is needed.

\section{Conclusion and Future Work}
Summarizing, we have presented an approach to lift a \clpfdbased\ solver to infinite domains.
Our approach tracks enumerations occurring during search and interprets how they affect the overall result.
Two extensions, high-level reasoning and random enumeration were used to increase applicability.

Techniques have been added to the animator and model checker \prob.
Here, they allowed us to use \prob\ as a prover and SMT solver.
Further applications in data validation show that \clpfd\ together with our extensions makes for a useful general purpose solver.

While we were pleasantly surprised by the overall performance, the extended \clpfd\ solver is still too weak to cope with the requirements a tool like \prob\ has.
Yet, it plays its part in combined solvers such as the one we outlined in~\cite{probz3integration} or in solver portfolios.

\bibliographystyle{eptcs}
\bibliography{infinite_domains}

\end{document}